\documentclass[10pt]{article}
\usepackage[OE]{express}
\usepackage{latexsym}
\usepackage{braket}
\usepackage{makeidx}
\usepackage{graphicx,epstopdf,subfigure,color}
\usepackage{hyperref}
\usepackage{physics} 

\def\vf#1{{[#1]}}
\def\va{\vf{\alpha}}

\def\vd{\vf{\alpha}}
\def\vsq{\vf{\beta}}
\def\vr{\vf{\theta}}
\def\vac{\ket{0}}
\def\pu{P_\uparrow}
\def\pk{P_{k-1,k}}

\makeindex

\begin{document}
\title{Dynamically reconfigurable sources for arbitrary gaussian states in integrated photonics circuits}

\author{Aharon Brodutch,\authormark{1,2,3,*}Ryan Marchildon,\authormark{1}and Amr S. Helmy\authormark{1,2}}

\address{\authormark{1}The Edward S. Rogers Department of Electrical and Computer Engineering, University of Toronto, 10 King's College Road, Toronto, Ontario M5S 3G4, Canada

\authormark{2}Center for Quantum Information and Quantum Control,  60 St George St., Toronto, Ontrario, M5S 1A7, Canada

\authormark{3}Department of Physics, University of Toronto, 60 St George St., Toronto, Ontrario, M5S 1A7, Canada}

\email{\authormark{*} brodutch@physics.utoronto.ca}

\begin{abstract}
We present a modular  design for integrated programmable multimode sources of arbitrary Gaussian states of light. The technique is based on current technologies, in particular recent  demonstrations of on-chip photon manipulation and the generation of highly squeezed vacuum states in semiconductors.  While the design is generic and independent of the choice of integrated  platform,  we adopt recent experimental results on compound semiconductors as a demonstrative example. Such a device would be valuable as a source for many quantum protocols that range from imaging to communication and information processing.
\end{abstract}





\section{Introduction}

\par Light provides an excellent platform for encoding quantum information that can be sent over long distances. In principle, the information encoded in light can be manipulated efficiently using currently available passive and active components, but many practical issues make the preparation and manipulation of such quantum states a difficult task in practice. While bulk optics provides a low-loss platform  to manipulate information encoded in small photonic systems, scalability remains a major problem. Integrated optics offers an effective route to mitigate scalability challenges, and several demonstrations of state preparation and control in  integrated optical devices have  been reported recently  \cite{Grafe2016,Tillmann2013,Carolan2015,Spagnolo2014}. These  were designed with linear optics quantum computing in mind, assuming that the information is encoded in finite dimensional systems using single photons.  In this work we show how to extend these schemes to the realm of continuous variable Gaussian states, by providing a blueprint for an integrated circuit that can be programmed and reconfigured to prepare any Gaussian state within a wide range of parameters. The design, approach and components utilized are based on currently available technologies, and rely on natural non-linearities in integrated waveguides to prepare initially squeezed vacuum states in multiple modes. 

Most quantum information processing (QIP) protocols have been designed for quantum systems with discrete degrees of freedom. These can be implemented using single photons with rail and/or polarization encoding \cite{KLMnature}. Such implementations suffer from several drawbacks such as the need for multiple synchronized single photon sources, photon-photon interactions that are difficult to achieve, gates that are probabilistic, and inefficient single-photon detection. Continuous variable (CV) QIP protocols that utilize light's continuous degrees of freedom offer several advantages over discrete approaches, in particular removing the requirement for single photons. In recent years both CV and hybrid CV/discrete approaches have been gaining interest as realistic approaches to  QIP \cite{Andersen2015, Weedbrook2012}. 

\par In CV protocols, the initial states are often Gaussian and can be generated from vacuum through a series of displacements, linear rotations, and squeezing  \cite{Ma1990}. Since these transformations are routinely achieved in bulk-optics, arbitrary Gaussian state generation seems straightforward in principle. In practice, however, the limited scalability and stability of bulk optics approaches is a hindrance to the development practical and large-scale quantum protocols, especially when the protocol must be scaled up to many modes, as in  CV cluster states \cite{Menicucci2006,Zhang2006} or the input states for some quantum simulations \cite{Huh2015}. Furthermore, the requirement for  in-line squeezing, i.e squeezing of an arbitrary state, is difficult even in bulk systems.

\par The ability to generate arbitrary multimode  Gaussian states  from an integrated chip would serve as an important milestone towards demonstrations of greater complexity and practical quantum technologies. Advances in the fabrication of integrated photonic circuits have made it possible to create large stable optical interferometers exhibiting low loss \cite{Harris_Nanophotonics_2016}. Moreover, semiconductor nonlinear waveguides have recently been used to produce squeezed vacuum states \cite{Dutt_OpticsLetters_2016, Dutt_PhysRevAppl_2015}. These components are sufficient  for generating and manipulating Gaussian light.

\par In this work we describe a generic architecture for integrated photonic devices that can be programmed to prepare arbitrary $N$-mode Gaussian states.  Our approach relies on a number of practical considerations: {\bf (1)}  It is much easier to generate squeezed vacuum states than to squeeze an arbitrary state. Consequently all squeezing is pushed to the beginning of the circuit. {\bf (2)} It is possible to modify Paris's approximate displacement method \cite{Paris1996} such that only a single displacement beam is needed, reducing the number of injected modes required from $\approx N$ to 2, regardless of $N$. {\bf (3)} It is possible to control all elements using tunable phase shifters. As a result the device can be fully programmable and dynamically reconfigurable using current technology. {\bf (4)} The design is modular, allowing easy adaptation to different material platforms and changing technologies, and is amenable to flip-chip implementations and hybrid-system approaches.  The result  allows a stable, programmable, scalable device that relies on current technological capabilities and can be implemented across a variety of photonic platforms. 

\section{Modular generation of  Gaussian states}\label{sec:modular1}

\par A state is called Gaussian if it has a Gaussian Wigner function; Equivalently, an $N$ mode state is a pure Gaussian state if and only if it can be generated from the vacuum by using a sequence of generalized multimode  \cite{Weedbrook2012,Schumaker1986} displacement $D(\vd)$, rotation $R(\vr)$ and squeezing $S(\vsq)$  operations, where the arguments are the multimode  \emph{displacement vector} $\vd$,  \emph{rotation matrix} $\vr$ and   \emph{squeezing matrix} $\vsq$ (see Sec. ~\ref{sec:methods} \ref{sec:Gaussian} for details). For Gaussian state generation it is possible to place the squeezing at the beginning of the sequence  \cite{Braunstein2005} and generate single-mode squeezed vacuum states in each mode (see Sec. ~\ref{sec:methods} \ref{sec:Gaussian} for the derivation)  removing the requirement for in-line  squeezing.

\par A realistic, modular approach to state generation can be based on the  decomposition  
\begin{equation} \label{eq:decompo} 
   \ket{G}=D(\va)R(\vf{\zeta})S(\vf{\beta^{1m}})\ket{0}
\end{equation}
where $\vf{\beta^{1m}}$ is a diagonal squeezing matrix indicating only single-mode squeezing.  Since any $N$ mode mixed Gaussian state can be created by tracing out $N$ modes from a $2N$ mode purification which is also a Gaussian state, we can restrict the discussion to pure states without loosing generality.

\subsection{Example: Generating  a pure  one-mode Gaussian state in bulk optics} \label{sec:singlemode}

\par
\begin{figure}[h!]
    \centering
    \includegraphics[width=0.8\columnwidth]{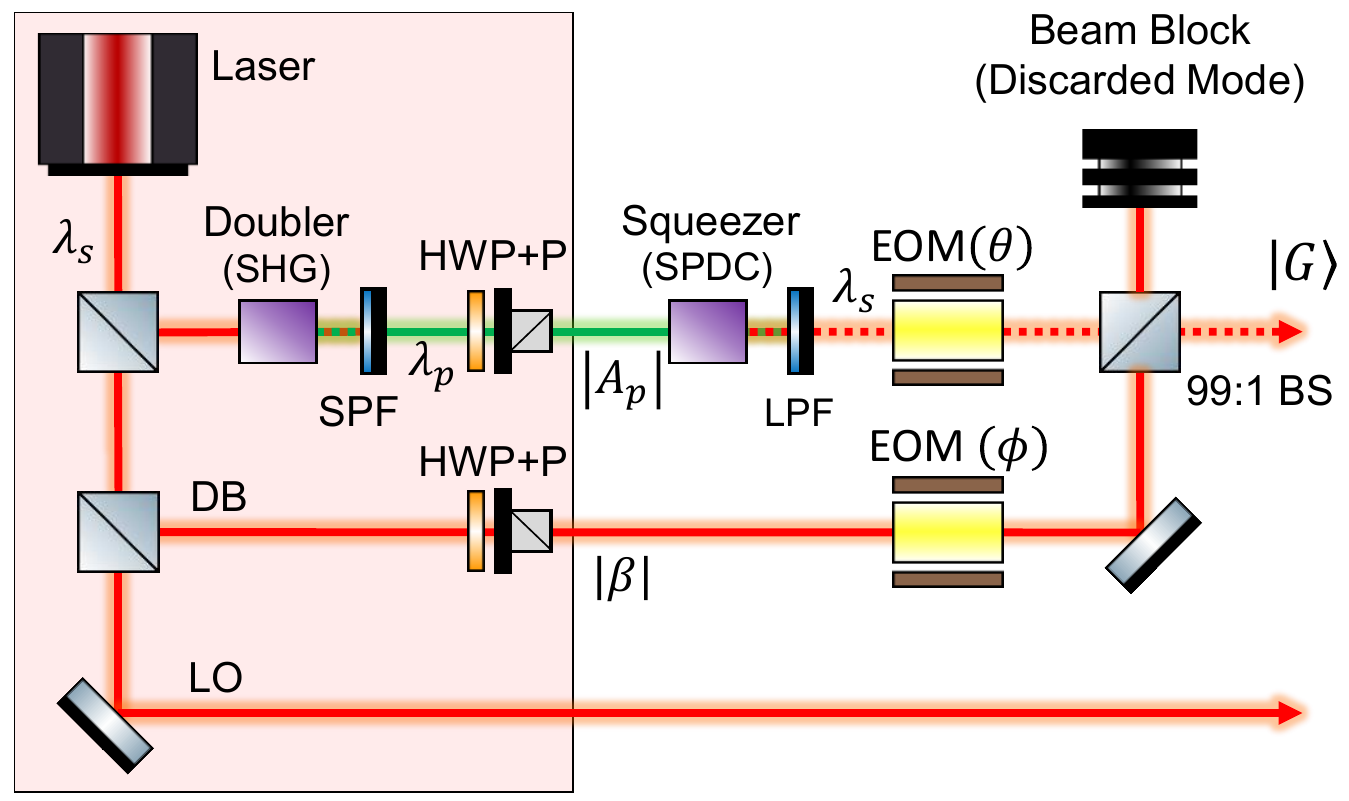}
     \caption{One-mode Gaussian state generation with bulk optics. The  initialization stage, highlighted in red, is also used in the $N$-mode protocol. Squeezed vacuum is generated through SPDC followed by a low-pass  filter (LPF) to remove the pump. An electro-optic modulator (EOM) is used to generate a phase shift. Displacement is generated via mixing with a high power phase-locked coherent state. Acronyms: HPF = high-pass filter; DB = displacement beam; LO = local oscillator; HWP+P = half wave plate and polarizer for amplitude control.}
    \label{fig:singleModeExample}
\end{figure}

\begin{figure}[t!]
	\centering
	\includegraphics[width=1\columnwidth]{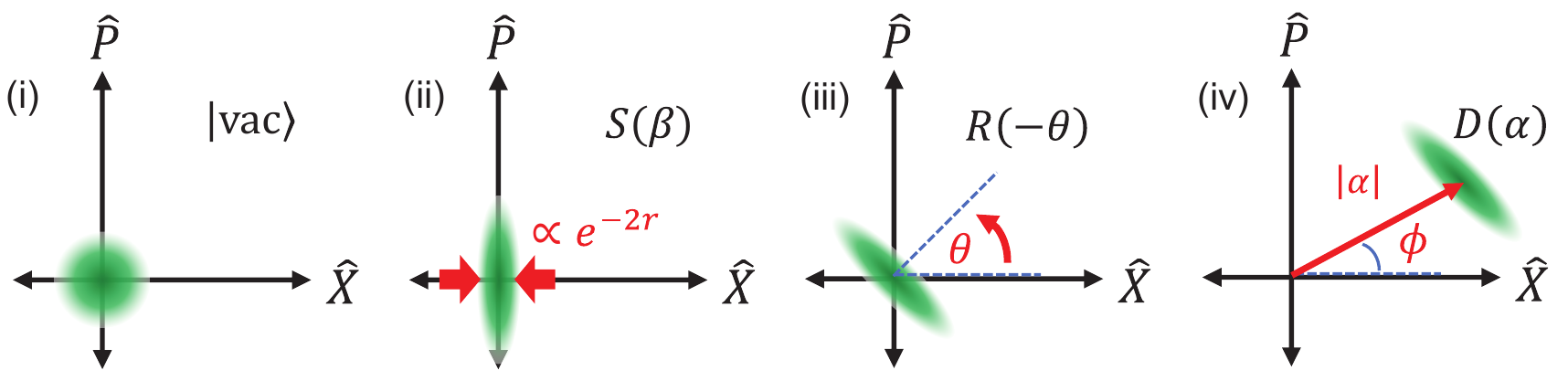}
	\caption{Phase-space depiction of  the four stage process. A one-mode vacuum state (i)  is squeezed (ii), rotated (iii) and displaced (iv).  }
	\label{fig:transformations}
\end{figure}

 To illustrate the applicability of the construction implied by Eq. \eqref{eq:decompo}, we describe  the generation of  pure one-mode  Gaussian states in bulk optics  as shown in Fig.~\ref{fig:singleModeExample}. The squeezing, rotation, and displacement transformations in phase space are depicted sequentially in Fig.~\ref{fig:transformations}. The scheme is as follows:

\par \textbf{(i) Initialization:} The protocol requires 2 phase locked beams, a signal wavelength $\lambda_{s}$ (e.g. 1550~nm) for displacement and a pump wavelength $\lambda_{p}$ (e.g. 775~nm) for generating the squeezed vacuum. In general it is useful to have an additional phase-locked beam of wavelength $\lambda_{s}$ to use as a reference or local oscillator (LO) for subsequent homodyne detection. A common approach is to coherently split a high intensity beam at $\lambda_{s}$ three ways, with one beam used to generate the $\lambda_{p}$ pump via second harmonic generation (SHG) in a nonlinear crystal `doubler' (e.g. BiBO). The squeezed light generated at  $\lambda_{s}$  (see below) maintains a stable  phase  relative to the other two beams.

\par \textbf{(ii) Squeezed vacuum preparation:}  Squeezing  (see Fig.~\ref{fig:transformations} ii)  through  spontaneous parametric downconversion (SPDC) \cite{Wu_PRL_1986} can be achieved using  a periodically-poled lithium niobate (PPLN) waveguide designed for squeezed light generation in the telecom C-band at $\lambda_{s}=\textrm{1550~nm}$ with a pump field at $\lambda_{p}=\textrm{775~nm}$. When operated in the single-pass configuration without optical feedback from a cavity, the output state has a squeezing parameter $r \propto \chi^{(2)}_{eff} \left\vert A_{\textit{p}} \right\vert L$, where $\chi^{(2)}_{eff}$ is the effective nonlinearity,  $\left\vert A_{\textit{p}} \right\vert$ is the pump amplitude, and $L$ is the waveguide length \cite{Lvovsky_arXiv_2016}. The  quadratures defined as  $\hat{x}=(\hat{a} + \hat{a}^{\dagger})/\sqrt{2}$ and $\hat{p}=(\hat{a}-\hat{a}^{\dagger})/i\sqrt{2}$,  have variances $\langle \Delta \hat{X}^{2} \rangle = e^{-2r}/2$ and $\langle \Delta \hat{P}^{2} \rangle = e^{2r}/2$.   The upper bound on $r$ is set by the parametric gain of the nonlinear medium, determined by both practical and physical limitations on $A_{p}$, $L$, and $\chi^{(2)}_{eff}$. The largest $r$ directly measured in a squeezed state to date is $r=1.73$ \cite{Vahlbruch_PhysRevLett_2016}, equivalent to 15~dB below the classical shot noise level.

\par \textbf{(iii) Rotation}: Arbitrary single-mode rotations (Fig.~\ref{fig:transformations} iii) require  a single phase-shifter to modify the  phase (relative to the reference). Note that by convention, rotations are defined in a clockwise  sense relative to the quadrature axes.

\par \textbf{(iv) Displacement:} The displacement operator $D(\alpha)$ (see Fig. \ref{fig:transformations} iv) can be approximated by mixing the squeezed state with a bright coherent state $\vert \alpha_0 \rangle$ at a  beamsplitter (or equivalent mode coupling device)  \cite{Paris1996} with reflectivity $\eta<<1$ such that $\sqrt{\eta}\alpha_0=\alpha$ (see Sec~\ref{sec:methods} \ref{sec:disp} for details). The deviation from an ideal displacement is plotted Fig.~\ref{fig:displacementScheme}(b).

\section{Generation of arbitrary $N$-mode Gaussian states on-chip}\label{sec:nmode}

\par The construction implied by \eqref{eq:decompo} can be used as a basis for a tunable on-chip $N$ mode Gaussian state generator. Below we describe a generic  approach for building such a device.  We then move to the simplest non-trivial example, a tunable two mode pure Gaussian state generator, using AlGaAs as a model platform. 

\subsection{Generic on-chip modular approach}\label{sec:arbitrary}

\par Recent work demonstrating on-chip squeezing \cite{Dutt_OpticsLetters_2016, Stefszky_PhysRevAppl_2017,Silverstone2013}, tunable phase-shifts \cite{Wang_OpticsCommunications_2014, Shadbolt_NaturePhoton_2012}, and arbitrary linear optical transformations \cite{Harris_Nanophotonics_2016} have assembled all the key ingredients necessary for independent control of $S(\vf{\beta^{1m}})$, $R(\vf{\theta})$, and $D(\va)$ in an integrated quantum circuit. In the generic setting, we consider each of the four stages as an independent module that can be fitted to the desired platform. Our aim is to show that the technology for each module has already been demonstrated at 1550~nm and to suggest possible implementations.

\par \textbf{Initialization}: The initialization stage consists of preparing coherent beams with stable relative phase: One at  775~nm  (pump) for use as a source for SPDC and one at  1550~nm for displacement.   Since  many QIP tasks would require coherence with an external reference, we consider the same external pumping as in the single-mode case of Sec. \ref{sec:modular1} \ref{sec:singlemode} (Fig.~\ref{fig:singleModeExample}). The external (bulk optics) stage does not affect scalability since the architecture suggested requires two input ports  regardless of $N$ (see Fig~\ref{fig:twomodeexample}).  We note that, a nonlinear waveguide that is phase-matched for SPDC with 775~nm can also achieve the 1550-to-775~nm SHG. However, this is limited by the amount of optical power that can be  handled by the chip without burning facets or inducing unwanted nonlinearities and does not offer an advantage in terms of scalability.

\par \textbf{Generation of squeezed vacuum:} 
 The key requirements for an on-chip source of strong squeezing are a high effective nonlinearity, low optical loss, and a long interaction length that is typically facilitated by the use of cavities due to limitations on the pump power. An on-chip squeezed light source based on low-loss silicon nitride microrings was recently demonstrated \cite{Dutt_OpticsLetters_2016}, where by controlling integrated microheaters to modify the cavity coupling, the measured squeezing was electrically tunable between 0.5~dB and 2~dB (corresponding to 0.9~dB and 3.9~dB when corrected for losses). This approach utilized a four-wave-mixing process stemming from third-order nonlinearities. In another approach, which utilized parametric downconversion from second-order ($\chi^{(2)}$) nonlinearities in a periodically-poled lithium niobate waveguide resonator \cite{Stefszky_PhysRevAppl_2017}, 2.9~dB of squeezing (corresponding to 4.9~dB in the lossless case) was directly measured. Both of these examples used continuous-wave pumping. Utilizing the higher peak powers and hence larger effective nonlinearities available through femtosecond pulsing can offer even higher degrees of squeezing. In Sec.~\ref{sec:example2m}, we describe an architecture based on AlGaAs that may be capable of producing squeezing in excess of 10~dB in a single-pass configuration under fs-pumping. 
 
 Programmable Mach Zehnder interferometers (MZIs) can be used to tune the squeezing parameter by attenuating the pump (see Fig.~\ref{fig:twomodeexample} and Sec. \ref{sec:example2m}).  Once squeezed light has been generated, the pump is typically filtered to prevent it from causing unwanted nonlinear phase modulation or squeezing elsewhere in the circuit. Common methods of filtering are wavelength demultiplexers built from asymetric coupled waveguides \cite{Guo_LSA_2017}, ring resonator filters, or Bragg reflector gratings \cite{Harris_PhysRevX_2014} which utilize the photonic bandgap effect.

\par \textbf{Rotation}: A generic rotation $R(\va)$ can be generated efficiently using an an array of linear optics elements (beam splitters and phase shifters) \cite{Reck1994,Clements_Optica_2016}.  In integrated circuits, tunable low-loss phase shifters can be achieved thermo-optically using resistive heaters to modify the local refractive index \cite{Shadbolt_NaturePhoton_2012}, or electro-optically using bias voltages \cite{Wang_OpticsCommunications_2014}, where the optimum choice depends on the material system. For example, AlGaAs circuits benefit from a strong electro-optic Pockel effect owing to a large intrinsic $\chi^{(2)}$ nonlinearity \cite{Wang_OpticsCommunications_2014}, whereas silicon-on-insultor (SOI) circuits possesses a relatively strong thermo-optic effect \cite{Harris_Nanophotonics_2016}. Beam-splitting transformations can be provided by directional couplers which evanescently couple optical modes between adjacent waveguides \cite{Yariv_JQuantElectron_1973}, or  multi-mode interferometers (MMIs) which work based on self-imaging effects \cite{Soldano_JLightTech_1995}. 3D multiport splitters can also be realized on-chip \cite{Spagnolo_NatureComms_2013}, but planar nearest-neighbour coupling remains the most compatible with conventional fabrication techniques. In-situ tunability over the splitting ratio is commonly achieved by concatenating a pair of two-mode splitters with a tunable phase shifter in one path between them, forming an MZI \cite{Wang_OpticsCommunications_2014}. An MZI with a tunable internal phase $\phi$ to control its splitting ratio, followed by an additional external phase shift $\theta$ in one outgoing arm, becomes the basic unit cell of reprogrammable circuits for universal rotations (see Fig.~\ref{fig:twomodeexample}). Recently, Harris et al. demonstrated a reprogrammable SOI quantum photonic chip comprised of 56 MZIs and 213 phase shifters \cite{Harris_Nanophotonics_2016}.

\par
\begin{figure}[h!]
    \centering
    \includegraphics[width=0.95\columnwidth]{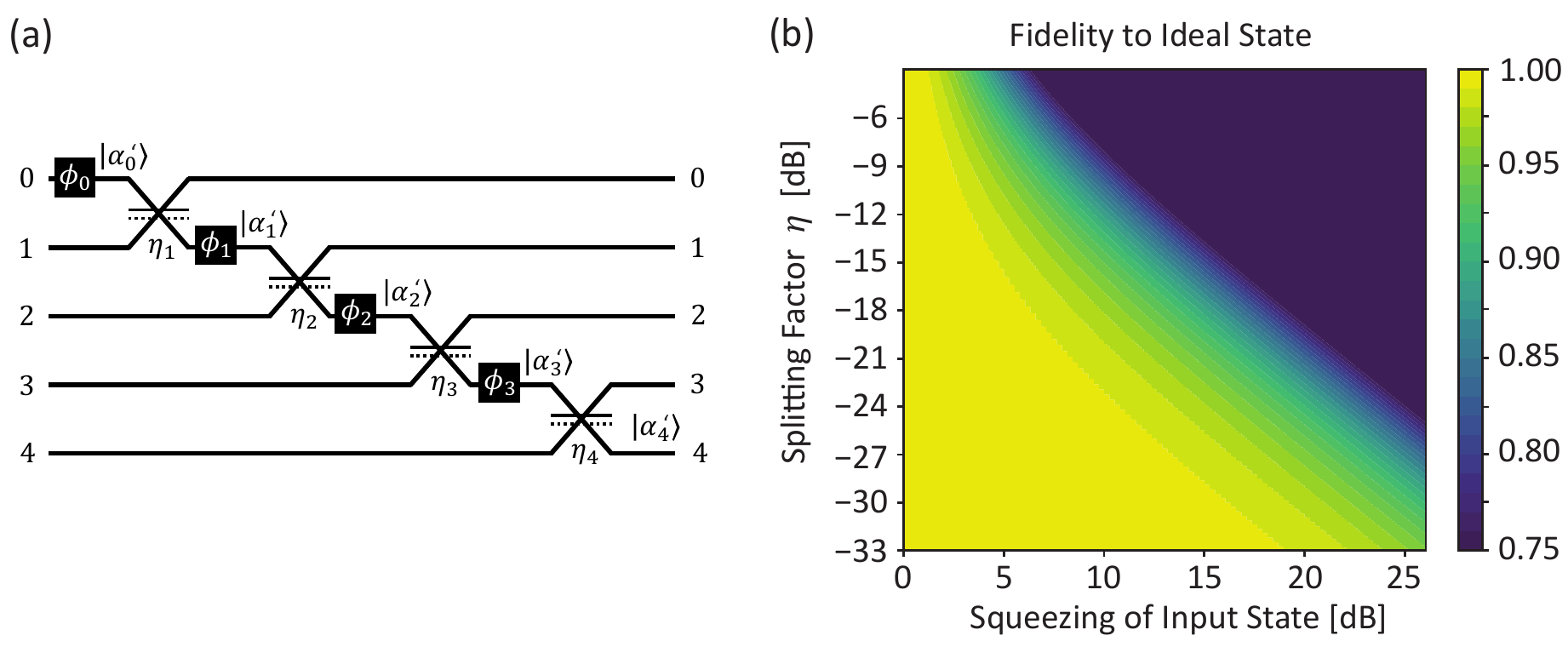}
     \caption{(a) The cascaded displacement scheme for $N=4$. At each step, a strong coherent state $\ket{\alpha_k'}$, in mode $k$ is used to displace mode $k+1$ by $\alpha_{k+1}=\alpha_k'\sqrt{\eta_k}$ and  swap with that mode using a beam splitter with reflectivity $\eta_k<<1$.   Note that the final modes are shifted by $-1$ with respect to the original modes so that the final mode $N$ is the ancilla which is discarded. 
      (b) Fidelity between a single-mode squeezed vacuum state after an approximate displacement $\alpha=0.5$ and the corresponding (ideal) displaced squeezed state (see Sec.~\ref{sec:boundsdisp}). Larger splitting factors $\eta$ increase the probability that photons from the squeezed state `leak' into the ancilla mode, or equivalently that noise is added by the ancilla, which degrades the squeezing and hence fidelity.}
    \label{fig:displacementScheme}
\end{figure}

 \textbf{Displacement:} It is possible to use Paris's method \cite{Paris1996} for approximating the displacement operator $D(\va)$  by pairing each mode with an ancillary strong coherent state mode and displacing each mode individually. However, such an architecture would be cumbersome to engineer with 2D planar waveguides and makes inefficient use of chip real-estate. Instead it is possible to use a single ancilla mode (mode $0$) containing a strong coherent state $\ket{\alpha_0}$ that cascades through an array of strongly cross-coupling mode splitters, displacing each mode sequentially as depicted in Fig.~\ref{fig:displacementScheme}(a). This  can be written as a rotation $R([\Delta])=\prod_k T_k$,  where $T_k$ is a  two-mode splitter transformation between modes  $k$ and $k-1$ with splitting factor (or reflectivity) $\eta_k$. 

\par At each coupling intersection, the strong coherent beam in mode $k-1$ displaces the state in mode $k$ and then the modes swap, i.e $\eta_k << 1$ and most of the light from mode $k$ is transmitted to $k-1$ and vice versa. If $\ket{\Psi}=R(\vf{\zeta})S(\vf{\beta^{1m}})\ket{0}$ is the state before the displacement, the approximate transformation can be written as (see Sec. \ref{sec:methods} \ref{sec:disp} for details):
\begin{equation}
    R([\Delta]) D_{0}(\alpha_{0}) \vert \Psi \rangle \otimes \vert \mathrm{vac} \rangle_{0}     \approx D_{N}(\alpha_{N})P_{\uparrow}\prod_{k=1}^{N}D_{k}(\alpha_{k}) \vert \Psi \rangle \otimes \vert \mathrm{vac} \rangle_{0} 
    \label{eqn:nModeDisplacementTransformation}
\end{equation}
where  $P_{\uparrow}$ is a permutation of modes that takes $0 \rightarrow N$ and $k \rightarrow k-1$ for all $k \in \{1, \cdots, N \}$ as in Fig. \ref{fig:displacementScheme}(a).

\par The displacement of each individual mode can be controlled by tuning the splitting factor $\eta_{k}$ of each mode coupler and rotating the phase, while taking into account all $\eta_{m}$ and $\alpha_{m}$ for which $m<k$. Tunability in $\eta_{k}$ can be achieved by implementing the mode coupler as an MZI with phase control \cite{Wang_OpticsCommunications_2014, Harris_Nanophotonics_2016}, or through  electro-optically or thermally inducing a modal mismatch between the two coupled waveguides \cite{Orlandi_OL_2013}. Adding phase shifters $\phi_{k}$ between stages to tune the phase of each $\vert \alpha_{k} \rangle$ allows control over the displacement angle. 

\par The first correction to the approximate displacement comes from the possibility that some photons from the displaced mode will leak into the displacement beam (see Sec.~\ref{sec:methods}~\ref{sec:disp}). Experimentally it is possible to put bounds on this error by blocking the displacement beam and counting the number of photons exiting port $N$. In general, the approximation will not be a dominant source of error as long as  $\eta_k$ is small compared to the probability that a single mode will lose a photon elsewhere in the circuit. In Sec.~\ref{sec:boundsdisp} (see also Fig.~\ref{fig:displacementScheme}(b)) we give a numerical example of the bounds on this approximation in the singe-mode case.

\subsection{Example: Arbitrary two-mode Guassian states generated in an AlGaAs integrated circuit}\label{sec:example2m}
 
\par
\begin{figure*}[h!]
    \centering
    \includegraphics[width=0.82\columnwidth]{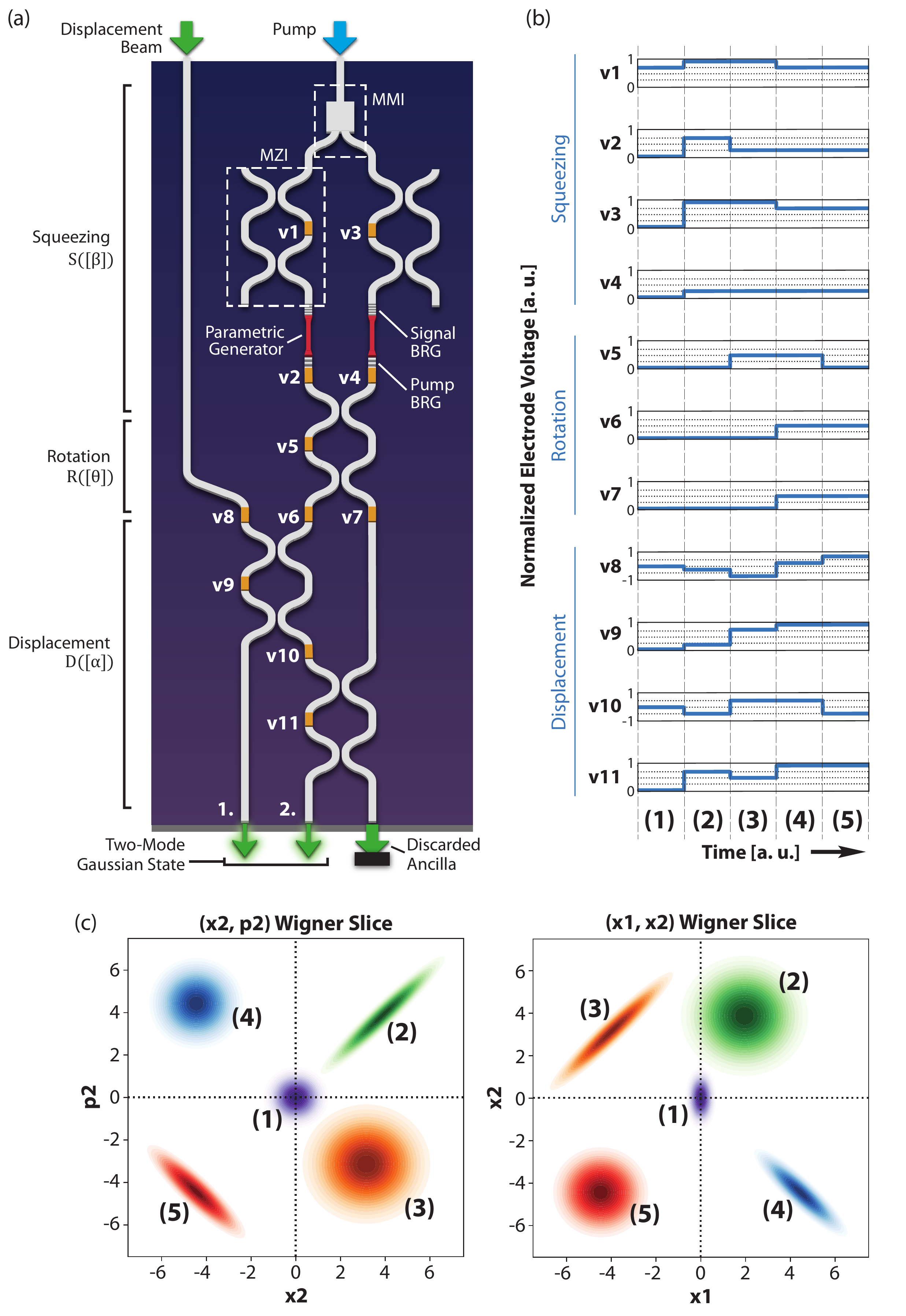}
     \caption{(a) Schematic of a dynamically reconfigurable  circuit for generating  two-mode  Gaussian states. The device includes the three reconfigurable modules (squeezing, rotation and displacement) fed by an external initialization module  as in Fig. \ref{fig:singleModeExample}.  The 11 electrodes can be used to program the state (see text, Sec \ref{sec:nmode}. \ref{sec:example2m}). Five example settings (b) generate the five states depicted in (c) where the (x2,~p2) Wigner slice shows the quadrature evolution in mode~2, while the (x1,~x2) slice shows correlations between modes. The states are: (1) squeezed vacuum in mode~1 and vacuum in mode~2; (2) two single-mode squeezed states (with mode~1 amplitude-squeezed and mode~2 phase-squeezed); (3) a displaced two-mode squeezed state where x1 and x2 are correlated; (4) a two-mode squeezed state with a different displacement amplitude and where x1 and x2 are now anti-correlated; and (5) two single-mode squeezed states rotated with respect to (2).} 
    \label{fig:twomodeexample}
\end{figure*}

 To give a concrete example of a device that includes all the elements above we describe a pure two-mode Gaussian state generator based on AlGaAs, as in Fig. \ref{fig:twomodeexample}(a).  We chose   AlGaAs  since it offers a broad range of quantum-circuit functionalities, including electro-optic tuning, self-pumped electrically-injected quantum state generation, and on-chip single photon detection \cite{Wang_OpticsCommunications_2014, Bijlani_APL_2013, Boitier_PRL_2014, Sprengers_APL_2011} (the latter two could become useful in various extensions of the device, for example generation of non-Gaussian states). It also supports a large intrinsic $\chi^{(2)}$  nonlinearity that facilitates the generation of squeezed vaccum states, with results indicating  squeezing parameters of $r>3$ in AlGaAs waveguides \cite{Zhizhongunpublished}. Here we consider the degenerate Type I parametric process where the downconverted photons are identical in frequency, polarization, and spatial mode. 

\par The circuit layout is shown in Fig.~\ref{fig:twomodeexample}(a), where eleven electrodes provide dynamic reconfigurability through electro-optic phase shifts, and MZIs serve as variable beamsplitters. To split the injected pump equally between the two parametric generator paths, we use a 1-by-2 port MMI due to the robustness of its fixed 50:50 splitting ratio against fabrication imperfections which eliminates the need for additional electrodes. Electrodes v1 and v3 adjust the fraction of pump power injected into the parametric generators, thereby tuning the squeezing, with v2 and v4 providing phase control. A Bragg reflection grating (BRG) filter blocking  1550~nm  is used to define the start of the parametric generator, while a second BRG blocking the $775$~nm pump terminates it. The parametric generator is a segment of the nonlinear waveguide that is narrowed in width. The narrowing adjusts the modal dispersion of the waveguide such that phase-matching is satisfied for 775~nm only within the narrowed segment, with the phase-matching tuning curve (e.g. see Fig.~2 in\cite{Horn_SciRep_2013}) shifting to shorter pump wavelengths as the waveguide width is increased \cite{Abolghasem_OL_2009}. Together with the BRGs, this provides a strategy for restricting squeezed light generation to only the desired region. Arbitrary rotations are provided via electrodes v5-v7. Finally, displacements are controlled by electrodes v8-v11  (as in Fig. ~\ref{fig:displacementScheme}(a)), where the MZIs are operated near conditions of perfect cross-coupling (mode swapping). 

 Fig.~\ref{fig:twomodeexample}(b) depicts the electrode voltages and corresponding output states for five different configurations. For readability we have re-normalized the voltage values to the following mappings: for squeezing $\textrm{v1}, \textrm{v3} \in [0,\,\, 1] \rightarrow r \in [0,\,\, r_{\textrm{max}}]$, with  $r_{\textrm{max}} = 1.0$ (8.7~dB); for single-mode phase rotations $\textrm{v2}, \textrm{v4}, \textrm{v6}, \textrm{v7} \in [0,\,\, 1] \rightarrow \theta \in [0,\,\, \pi]$; for two-mode mixing $\textrm{v5} \in [0,\,\, 1] \rightarrow \eta \in [1,\,\, 0]$; for displacement angle $\textrm{v8, v10}\in[-1,\,\, 1] \rightarrow \phi \in [-\pi,\,\, \pi]$; and for displacement magnitude $\textrm{v9}, \textrm{v11} \in [0,\,\, 1] \rightarrow \eta \in [0,\,\, 0.0125]$, where the resultant displacement of mode $k$ is $D_{k}(\alpha_{k}' \sqrt{\eta_{k+1}})$ (see Fig.~\ref{fig:displacementScheme}(a)) and  $\left\vert \alpha_{0} \right\vert = 40$. Note that we remain under the estimated bound of $\eta \leq 0.014~(-18.4~\mathrm{dB})$ needed to maintain fidelities of 98\% or greater with the ideal displaced state (see Sec.~\ref{sec:disp}). Fig.~\ref{fig:twomodeexample}(c) shows two Wigner function slices from the output state, computed for five different configurations after tracing out the ancilla mode.

\par The phase shifters envisaged in Fig.~\ref{fig:twomodeexample} are based on electro-optic modulation as seen in previous AlGaAs quantum circuits \cite{Wang_OpticsCommunications_2014}. Circuit reconfigurability can be achieved using a myriad of techniques, some being more favourable than others depending on the specific needs of the application. Electro-optic and thermal tuners have the advantage of being implementable monolithically on the same platform as passive components, with the former capable of achieving modulation speeds in the GHz, while the latter is limited by the thermal time constant but can achieve switching speeds in the MHz when appropriately designed. In cases where performance enhancements such as higher speed, better switching extinction, or lower bias voltages are needed, advanced coupler designs such as grating-assisted, asymmetric, or ring-resonator couplers can be used at the expense of a reduced operating bandwidth \cite{Erdogan_OC_1998, Wang_OL_2007, Manipatruni_OE_2010}. Whereas a simple electro-optic MZI coupler may need up to tens of volts of bias, a ring resonator coupler can require merely a fraction of a volt. In some cases, flip-chip bonding with active devices may be appropriate, but this comes at the cost of increased optical loss, and hence is only really suitable for modulation of the pump. For example, rapid tuning of the squeezing parameter $r$ can be achieved with speeds exceeding 10~GHz via absorption-based modulation of the pump using the quantum-confined stark effect \cite{Miller_PRL_1984} with flip-chip bonded III-V semiconductors.

\section{Methods}\label{sec:methods}
\subsection{Gaussian states and unitaries} \label{sec:Gaussian} 

An $N$ mode Gaussian unitary operation can be decomposed into an $N$ mode rotation followed by $N$ mode squeezing followed by $N$ mode displacement \cite{Cariolaro2016}
\begin{equation}\label{eq:Gaussian:generic}
U_G(\vr,\vsq,\va)=D(\va)S(\vsq)R(\vr),\end{equation}
where the displacement vector $\vd$ has $N$ complex entries $\alpha_k$ representing the displacement of each mode; the rotation matrix $\vr$ is an $N\times N$ unitary matrix with entries $\theta_{k,l}$; and the squeezing matrix $\vsq$ is a  complex, symmetric $N\times N$ matrix with entries $\beta_{k,l}$. These operators can be written in Fock notation as:

\begin{itemize}
	\item {\it Displacement:} $D(\vd)=\exp{\sum_k\left(\alpha_ka^\dagger_k-\alpha^*ka_k\right) }$,  
	\item {\it Rotation:} $R(\vr)=\exp{i\sum_{k,l}\left( \theta_{k,l}a^\dagger_ka_l\right)}$, and 
	\item {\it Squeezing:} $S(\vsq)=\exp{\sum_{k,l}\left(\beta_{k,l}a^\dagger_ka^\dagger_l-\beta_{k,l}^*a_ka_l\right)}$.
\end{itemize}
 
A pure Gaussian state $\ket{G}$ is generated by $U_G(\vr,\vsq,\va)\vac$, however since a rotation at most adds a global phase to the vacuum state it is possible to remove the first rotation step: $\ket{G}=U_G(\vr=0,\vsq,\va)\ket{0}$.
To bring the squeezing matrix into diagonal form  $S(\vf{\beta^{1m}})$ (i.e a form that implies only one mode squeezing) we use the following 
facts (see \cite{Ma1990,Cariolaro2016}): 

\begin{itemize}

\item $R(\vf{\zeta}))S(\vf{\beta})=S(\vf{\beta'})R(\vf{\zeta})$ such that $\vf{\beta'}=e^{i\vf{\zeta}}\vf{\beta}{e^{i\vf{\zeta}}}^T$, where $T$ means transpose. 

\item It is possible to bring any symmetric matrix $\vf{\beta}$ into diagonal form using the Takagi factorization, i.e for any $\vf{\beta}$ there exists a unitary $U$ such that  $\vf{\beta^{1m}}=U\vf{\beta}U^T$ is a diagonal matrix with non-negative entries. 

\end{itemize}

It follows (see \cite{Braunstein2005} for details) that we can write  
\begin{equation}
    U_G(\vr,\vsq,\va)=D(\va)R(\vf{\zeta})S(\vf{\beta^{1m}})R(\vf{\zeta})R(\vr)\end{equation}
  
The above can be simplified further in the case of Gaussian states since $ R(\vf{\zeta})R(\vr)\ket{0}=\ket{0}$, which leads us to \eqref{eq:decompo}.

\subsection{Approximate displacement}\label{sec:disp}

The scheme in Fig \ref{fig:displacementScheme} generates an approximate displacement $D(\va)$ as well as a permutation of the modes, as long as  $\eta_k$ is small enough. It is easier to see how well the approximation of \eqref{eqn:nModeDisplacementTransformation} works by writing $T_k$ in exponential form: 
\[T_k=\pk e^{i\delta_{k-1,k}(a^\dagger_{k-1}a_k+a^\dagger_ka_{k-1})}e^{i(\phi_k-\phi_{k-1})a_{k-1}^\dagger a_{k-1}}\] 
with $\pk$ being the operator that swaps modes $k-1$ and $k$, $\phi_k$ is the phase of $\alpha_k$ and $\delta_{k-1,k}$ defined via  $\sqrt{\eta_k}=sin(\delta_{k-1,k})=\frac{\alpha_k}{\alpha_0\prod_{m=1}^{k-1}Cos(\delta_{m-1,m})}$ with  $\delta_{-1,0}=0$. Note that for simplicity of the derivation we neglect any phases added by the cross-coupling since these can always be corrected.  

Using the rules for switching the order of rotations and displacements, we can rewrite the transformation 
\begin{align}
    &\nonumber T_1D_{0}(\alpha_0) \\ \nonumber &= P_{1}D_0(e^{-\phi_1}\alpha_0 \cos\delta) D_1(\alpha_1)e^{i\delta_{0,1}(a^\dagger_{0}a_1+a^\dagger_1a_{0})}e^{-\phi_1a_0^\dagger a_0}\\ \nonumber&=D_0(\alpha_1)D_1(e^{-\phi_1}\alpha_0 \cos\delta) P_1e^{i\delta_{0,1}(a^\dagger_{0}a_1+a^\dagger_1a_{0})} e^{-\phi_1a_0^\dagger a_0}
\end{align}
which can be done for all terms, so that 
\begin{align}
    \prod_{k=1}^N T_kD_{0}(\alpha_0)=&D_{0..N}(\va)  \times\\&   D_N\left[e^{i\phi_N}\alpha_0 \prod_{k=N}^NCos(\delta_{k-1,k})\right]\prod_{k=1}^NT_k\nonumber
\end{align}
where we use the ordering convention $\Pi_{k=1}^NX_k=X_NX_{N-1}...X_1$. It is possible to move the permutations to the left and get 
\begin{equation}
    \prod_{k=1}^NT_k=\pu \prod_{k=1}^N e^{i\delta_{k-1,k}(a^\dagger_{0}a_k+a^\dagger_ka_{0})}         e^{i(\phi_N)a_{0}^\dagger a_{0}}
\end{equation}
Now taking $|\alpha_0|>>1$ such that $\delta_{k-1,k}<<1$ for all $k$ we can see that $\prod_{k=1}^NT_k R(\vr)S(\vf{\beta^{1m}})\vac\approx R(\vr)S(\vf{\beta^{1m}})\vac$. With first order corrections being  $\sum_k\delta_{k-1,k}(a^\dagger_Na_{k-1})R(\vr)S(\vf{\beta^{1m}})\vac$. These can be treated as possible photon losses.

\subsection{Simulation}\label{sec:simulation}
Numerical results were obtained using a simulation written in Python. The state of the system was encoded in a displacement vector $\vec{d}$ and covariance matrix $\vec{\sigma}$ which were updated at each optical element. The dynamical maps for closed system (i.e unitary) elements (squeezing, phase shifters and beam splitters) were represented using symplectic transformations as per standard methods ({e.g. see \cite{Adesso2014}}). Noise due to dissipation (for the results in Sec.~\ref{sec:loss}) was added using a standard transformation $\vec{d}\rightarrow \tau \vec{d}$, $\vec{\sigma} \rightarrow \tau \vec{\sigma} +(1-\tau)\vec{\sigma_V}$ where $\tau$ is the transmission of the lossy optical element and $\vec{\sigma_V}$ is the covariance matrix for the vacuum state \cite{Ferraro2005}. Two types of loss were included: single-mode loss added after phase shifters and squeezers, and two-mode loss added to each mode separately before and after each directional coupler or MZI.


\section{Challenges and limitations}

\label{sec:challanges}

\par  Throughout this work we assumed that everything is ideal and neglected the corrections due to the approximate displacement stage.  Below we discuss the main issues that will constrain performance.

\subsection{Bounds for single displacement stage} \label{sec:boundsdisp}

\par In principle, it should be possible to have arbitrarily good approximations of any displacement as long as $\alpha_0$ is large enough. In practice however, the larger we make $\alpha_0$, the more precisely we will need to control each $\eta_k$. It is therefore useful to fix $\alpha_0$ and establish an  upper bound on $\eta_k$ such that the fidelity with the ideal displacement is sufficiently high. We  consider the simple case of a single-mode squeezed vacuum state $\vert \psi\rangle$. The total state before displacement is given by $ \vert \Psi \rangle = \vert \alpha_{0} \rangle_{0} \otimes \vert \psi \rangle_{1}.$ In the ideal case  the output state is $
    \vert \Psi' \rangle_{\textrm{Ideal}} = D_{0}(\alpha_{0}\sqrt{\eta_{1}})R_{0}(\pi/2)\vert \psi\rangle_{0} \otimes R_{1}(\pi/2) \vert \alpha_{0} \sqrt{1-\eta_{1}} \rangle_{1},$
whereas applying the standard mode-mixing transformation to the state gives the true output: $
    \vert \Psi' \rangle_{\textrm{Actual}} = T_1\left[ \vert \alpha_{0} \rangle_{0} \otimes \vert \psi \rangle_{1}\right]$.
    
Fig. ~\ref{fig:displacementScheme}(b) shows the Uhlmann fidelity \cite{Uhlmann_RMP_1976} between $\vert \Psi'\rangle_{\textrm{Ideal}}$ and $\vert \Psi'\rangle_{\textrm{Actual}}$ computed as a function of $\eta$ and input state squeezing when $\ket{\psi}$ is a squeezed vacuum state and the displacement is fixed at $\alpha=0.5$.  As the squeezing increases, we see that high fidelity requires smaller $\eta$ (and larger $\alpha_0$). The bound on $\eta$ for obtaining a fidelity of at least $F$ is given in dB by $\eta_{\mathrm{[dB]}} \leq (a r^{-b} + c)$, where for $F\geq 95\%$ we have $\{a=-30.35, \, b=0.39, \, c=16.11\}$, and for $F \geq 98\%$ we have $\{a=-40.61,\, b=0.29, \, c=22.15 \}$. For squeezing of up to nearly $r=0.5$~(4.3~dB), $\eta$ can be kept relatively large at around -10 dB. However, for an input state with 15~dB of squeezing ($r=1.73$) we require $\eta \leq -21.5 \mathrm{dB}$ and $\eta \leq -25.5 \mathrm{dB}$ for $F \geq 95\%$ and $F \geq 98\%$ respectively. Though small, these reflectivities are within range of existing integrated photonics technologies; Harris et al. recently demonstrated tunable MZIs with extinction ratios of -66~dB \cite{Harris2017}.

\subsection{Squeezing}

\par Squeezing in the waveguide can be increased by either increasing the length of the squeezing stage, or by increasing the 775~nm beam power. In practice too much pump power can damage the device, have unwanted effects such as self-phase modulation, or be self-limiting through two-photon absorption which increases with the pump intensity. Increasing length has two problems: first it will require a larger device, but more significantly it will increase loss (see below). These will limit the maximum achievable squeezing per mode in a real device. One target to aim for is the $\sim 20.5$~dB needed for fault-tolerant cluster-based quantum computing using Gottesman-Kitaev-Preskill encoding \cite{Menicucci_PRL_2014}.

\subsection{Loss}\label{sec:loss}

\par Minimizing optical loss is crucial to fully benefit from the squeezing achievable in a given platform. This can prove quite challenging in practice owing to how quickly the squeezing decays as loss increases. The amount of measurable squeezing falls as $S_{T} = 10 \cdot \textrm{log}_{10}\left[T \cdot 10^{-S_{0}/10} + (1-T)\right]$ where $S_{0}$ and $S_{T}$ refer to the measurable squeezing in dB before and after losses respectively, and $T$ is the total transmission efficiency \cite{Lvovsky_arXiv_2016}. Optical losses in an integrated circuit can be caused by waveguide sidewall roughness, mode leakage at waveguide bends, reflections at material interfaces (such as the waveguide facets), or modal mismatches when coupling into and out of the devices.  Loss therefore poses the most problematic constraint for scalability, since for arbitrary rotations the device length grows quickly with the number of modes (around $n^2$ per the Reck scheme \cite{Reck1994}), and loss is exponential in the circuit length. 

\par
\begin{figure}[h!]
    \centering
    \includegraphics[width=0.95\columnwidth]{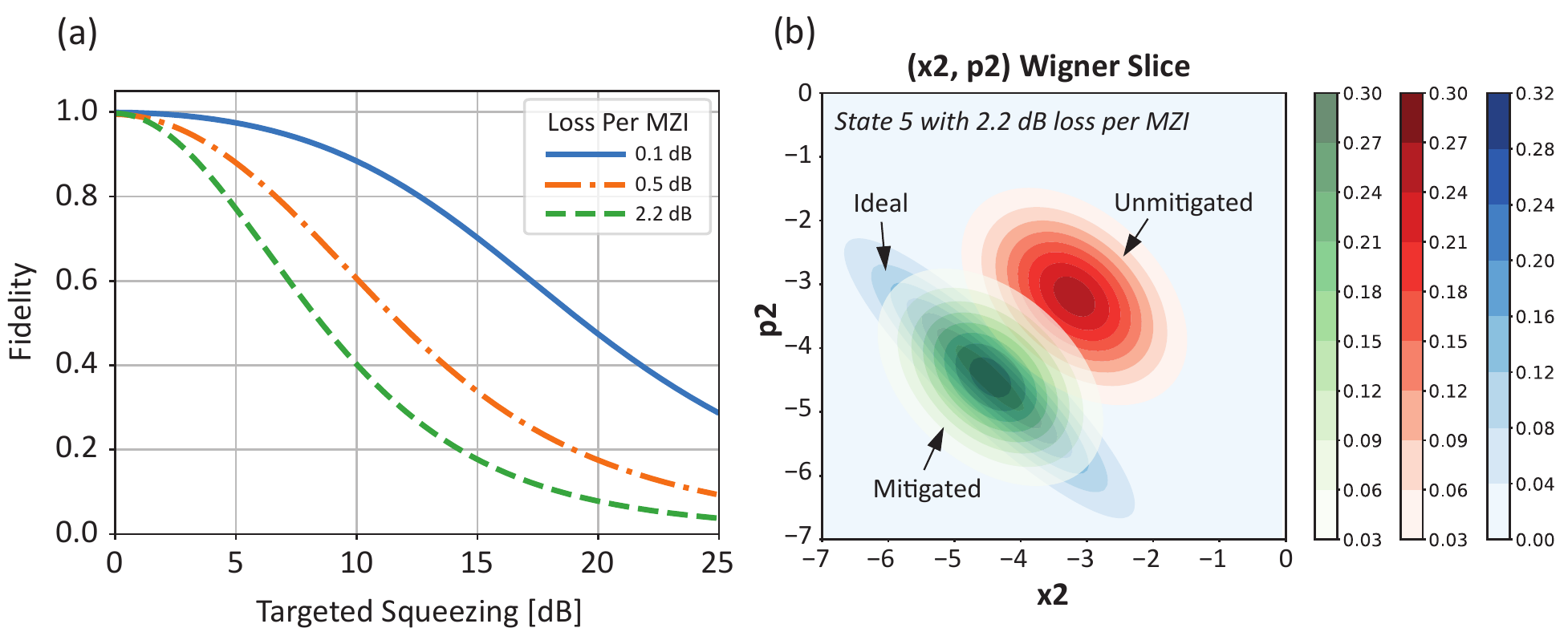}
    \caption{Impact of loss on the preparation of a squeezed two-mode state by the circuit in Figure~\ref{fig:twomodeexample}, with the displacement stage parameters tuned to mitigate errors in targeted vs. actual displacement. (a)~Uhlmann fidelity as a function of the squeezing parameter for various values of MZI loss (the loss per coupler and phase-shifter are taken as a fixed fraction of the MZI loss). (b)~Comparing the ideal preparation of State~5 in Fig.~\ref{fig:twomodeexample} to a lossy preparation, with and without mitigation via modification to the displacement stage parameters.}
   \label{fig:loss}
\end{figure}

 In Fig.~\ref{fig:loss}(a) we have simulated the impact of loss on the two-mode Gaussian states generated by the circuit in Fig.~\ref{fig:twomodeexample}, as a function of the targeted squeezing. Errors in displacement due to loss have been mitigated by small adjustments to electrode voltages v9 and v11, which compensate for attenuation of the displacement beam's amplitude $\alpha$ and subsequent losses after each displacement stage. The result of this strategy is illustrated in Fig.~\ref{fig:loss}(b), where the targeted state displacement has been restored. Our computed fidelity is therefore determined solely by the loss of correlations and squeezing in the output state. In Fig.~\ref{fig:loss}(a), the insertion loss of 2.2~dB/MZI is based closely on the AlGaAs circuit reported in Ref.~\cite{Wang_OpticsCommunications_2014}, where the waveguide propagation loss was 1.6~dB/cm. With further engineering, the length of each coupler and phase shifter could be decreased to help reduce this loss substantially. As a benchmark for low-loss quantum circuits, we consider the 0.1~dB/MZI achieved in Ref.~\cite{Harris_Nanophotonics_2016} for a silicon-on-insulator (SOI) platform. In this latter case, although propagation losses were higher at 2.4~dB/cm, the phase shifters and couplers were achieved far more compactly, at tens of microns length rather than several millimeters. Nonetheless, even for low losses of 0.1~dB/MZI, achieving high fidelity for states targeting high squeezing ($>$5~dB) remains challenging. 
 
\par One possible mitigation strategy is entanglement distillation, which uses local non-Gaussian elements (such as photon counting) and sacrificial ancilla states to enhance the purity and correlations of a state subjected to loss \cite{Andersen2015, Fiurasek_PRA_2011}. This can benefit from the relative ease in which a large number of ancillas can be prepared on an integrated chip compared to bulk approaches. Another distillation approach is to use heralded noiseless linear amplifiers \cite{Xiang_NaturePhoton_2010,Zavatta_NaturePhoton_2010}, which can be realized compactly in integrated optics, and in the case of AlGaAs, could even be monolithically embedded within the same platform. 

\section{Conclusions and outlook}

\par We provided a generic architecture for a device that can prepare arbitrary multimode Gaussian states. These states are useful for  various  tasks \cite{Weedbrook2012} including quantum sensing \cite{Tan2008}, quantum communication \cite{Niset2008,Hayden2016} and quantum computing \cite{Menicucci2006}. The device would be capable of generating the optimal states for these protocols, and will motivate further research on optimization of CV protocols under realistic constraints such as loss. The device is programmable, making it versatile and robust to some imperfections and bias drifts which can be actively corrected.  Fast reconfigurability would also aid  in correcting  optical phase drifts.  Given current technologies, the most significant limiting factor of such devices would be loss, which would both limit squeezing and modify the outgoing state. Reconfigurability allows some loss mitigation, in particular it is possible to compensate for attenuated displacement by modifying the displacement stage.  

Beyond Gaussian states, the device  provides a scalable route for generating more general CV states, for example by placing detectors at some of the outputs and post-selecting a desired non-Gaussian state or even using adaptive schemes. Since these may be difficult to simulate, it may be useful to optimize the control parameters via feedback as in recent  NMR protocols \cite{Lu2017a,Dive2017}. The scheme  can also be  modified to perform arbitrary multimode Gaussian  operations, however such a device will require in-line squeezing which is very demanding in practice.  Fast dynamical reconfigurability and appropriately placed time delays on some modes can also allow us to use the temporal degree of freedom  to encode larger states \cite{Yokoyama_NaturePhoton_2013, Yoshikawa_APLPhoton_2016}.

\par We showed that the device can be implemented  in AlGaAs, but kept the design generic and modular so that it would could be implemented using a variety of material systems which can be chosen based on their particular advantegers for the task at hand, be it reduced loss, faster reconfigurability, or integrability with active devices.   Since photonic  technology  for implementing such devices already exists, we expect to see a practical demonstration of our scheme in the near future. Such a demonstration would be a significant milestone for CV QIP. 

\section*{Funding}
 Center For Quantum Computation and Quantum Control, University of Toronto.

\section*{Acknowledgments}

 We thank Daniel James, Daniel Giovannini, Zhizhong Yan, Gleb Egorov and Armanpreet Pannu for discussions and advice.


\begin{thebibliography}{1}
\newcommand{\enquote}[1]{``#1''}

\bibitem{Grafe2016}
M.~Gr{\"{a}}fe, R.~Heilmann, M.~Lebugle, D.~Guzman-Silva, A.~Perez-Leija, and
  A.~Szameit, \enquote{{Integrated photonic quantum walks},} J.  Opt.
  \textbf{18}, 103002 (2016).

\bibitem{Tillmann2013}
M.~Tillmann, B.~Daki{\'{c}}, R.~Heilmann, S.~Nolte, A.~Szameit, and P.~Walther,
  \enquote{{Experimental boson sampling},} Nat. Photonics
 \textbf{7},
  540--544 (2013).

\bibitem{Carolan2015}
J.~Carolan, C.~Harrold, C.~Sparrow, E.~Mart{\'{i}}n-L{\'{o}}pez, N.~J. Russell,
  J.~W. Silverstone, P.~J. Shadbolt, N.~Matsuda, M.~Oguma, M.~Itoh, G.~D.
  Marshall, M.~G. Thompson, J.~C.~F. Matthews, T.~Hashimoto, J.~L. O'Brien, and
  A.~Laing, \enquote{{ Universal linear optics.}} Science \textbf{349}, 711--6 (2015).

\bibitem{Spagnolo2014}
N.~Spagnolo, C.~Vitelli, M.~Bentivegna, D.~J. Brod, A.~Crespi, F.~Flamini,
  S.~Giacomini, G.~Milani, R.~Ramponi, P.~Mataloni, R.~Osellame, E.~F.
  Galv{\~{a}}o, and F.~Sciarrino, \enquote{{Experimental validation of photonic boson sampling},} Nat. Photonics
 \textbf{8}, 615--620 (2014).

\bibitem{KLMnature}
E.~Knill, R.~Laflamme, and G.~Milburn, \enquote{{A scheme for efficient quantum
  computation with linear optics},} Nature \textbf{409}, 46--52 (2001).

\bibitem{Andersen2015}
U.~L. Andersen, J.~S. Neergaard-Nielsen, P.~van Loock, and A.~Furusawa,
  \enquote{{Hybrid discrete- and continuous-variable quantum information},}
  Nat. Physics \textbf{11}, 713--719 (2015).

\bibitem{Weedbrook2012}
C.~Weedbrook, S.~Pirandola, R.~Garc{\'{i}}a-Patr{\'{o}}n, N.~J. Cerf, T.~C. Ralph, J.~H. Shapiro, and S.~Lloyd, \enquote{{Gaussian quantum information},} Rev. Mod. Phys. \textbf{84}, 621--669 (2012).

\bibitem{Ma1990}
X.~Ma and W.~Rhodes, \enquote{{Multimode squeeze operators and squeezed
  states},} Phys. Rev.  A \textbf{41}, 4625--4631 (1990).

\bibitem{Menicucci2006}
N.~C. Menicucci, P.~van Loock, M.~Gu, C.~Weedbrook, T.~C. Ralph, and M.~A.
  Nielsen, \enquote{{Universal quantum computation with continuous-variable
  cluster states},} Phys. Rev.  Lett.  \textbf{97}, 110501 (2006).

\bibitem{Zhang2006}
J.~Zhang and S.~L. Braunstein, \enquote{{Continuous-variable Gaussian analog of
  cluster states},} Phys. Rev.  A \textbf{73}, 032318 (2006).
  
  \bibitem{Huh2015}
J.~Huh, G.~G. Guerreschi, B.~Peropadre, J.~R. McClean, and A.~Aspuru-Guzik,
  \enquote{{Boson sampling for molecular vibronic spectra},} Nat. Photonics  \textbf{9}, 615--620 (2015).

 \bibitem{Harris_Nanophotonics_2016} 
N.~Harris, D.~Bunandar, M.~Pant, G.~Steinbrecher, J.~Mower, M.~Prabhu, T.~Baehr-Jones, M.~Hochberg, and D.~Englund, \enquote{{Large-scale quantum photonic circuits in silicon},} Nanophotonics \textbf{5}, 456--468 (2016).

  

  
\bibitem{Dutt_OpticsLetters_2016}
A. Dutt, S. Miller, K. Luke, J. Cardenas, A. Gaeta, P. Nussenzveig, and M. Lipson, \enquote{{Tunable squeezing using coupled ring resonators
on a silicon nitride chip},} Opt. Lett. \textbf{41}, 223--226 (2016).

\bibitem{Dutt_PhysRevAppl_2015}
A. Dutt, K. Luke, S. Manipatruni, A. Gaeta, P. Nussenzveig, and M. Lipson, \enquote{{On-chip optical squeezing},} Phys. Rev.  Applied \textbf{3}, 044005 (2015). 

\bibitem{Paris1996}
M.~G. Paris, \enquote{{Displacement operator by beam splitter},} Phys. 
  Lett. A \textbf{217}, 78--80 (1996).

 \bibitem{Schumaker1986}
B.~L. Schumaker, \enquote{{Quantum mechanical pure states with Gaussian wave
  functions},} Phys. Rep. \textbf{135}, 317-408  (1986).
  
  \bibitem{Braunstein2005}
S.~L. Braunstein, \enquote{{Squeezing as an irreducible resource},} Phys. Rev.  A \textbf{71}, 055801 (2005).


\bibitem{Wu_PRL_1986}
L. Wu, H. Kimble, J. Hall, and H. Wu, \enquote{{Generation of squeezed states by parametric down conversion},} Phys. Rev.  Letters \textbf{57}, 2520 (1986). 



\bibitem{Lvovsky_arXiv_2016}
A. Lvovsky, \enquote{{Squeezed Light},} in {\it Photonics, Volume 1: Fundamentals of photonics and physics, Ed. D. Andrews}  (2015).


\bibitem{Vahlbruch_PhysRevLett_2016} H. Vahlbruch, M. Mehmet, K. Danzmann, and R. Schnabel, \enquote{{Detection of 15 dB squeezed states of light and their application for the absolute calibration of photo-electric quantum efficiency},} Phys. Rev.  Lett.  \textbf{117}, 110801 (2016). 


\bibitem{Silverstone2013}
J.~W. Silverstone, D.~Bonneau, K.~Ohira, N.~Suzuki, H.~Yoshida, N.~Iizuka,
  M.~Ezaki, C.~M. Natarajan, M.~G. Tanner, R.~H. Hadfield, V.~Zwiller, G.~D.
  Marshall, J.~G. Rarity, J.~L. O'Brien, and M.~G. Thompson, \enquote{{On-chip
  quantum interference between silicon photon-pair sources},} Nat. Photonics
  \textbf{8}, 104--108 (2013).

\bibitem{Stefszky_PhysRevAppl_2017}
M. Stefszky, R. Ricken, C. Eigner, V. Quiring, H. Herrmann, and C. Silberhorn, \enquote{{Waveguide cavity resonator as a source of optical squeezing},} Phys. Rev.  Applied \textbf{7}, 044026 (2017). 


\bibitem{Shadbolt_NaturePhoton_2012}
P. Shadbolt, M. Verde, A. Peruzzo, A. Politi, A. Laing, M. Lobino, J. Matthews, M. Thompson, and J. O'Brien, \enquote{{Generating, manipulating and measuring entanglement and mixture with a reconfigurable photonic circuit},} Nat. Photonics
 \textbf{6}, 45 (2012)

\bibitem{Wang_OpticsCommunications_2014}
J.~Wang, A.~Santamato, P.~Jiang, D.~Bonneau, E.~Engin, J.~Silverstone, M.~Lermer, J.~Beetz, M.~Kamp, S.~Höfling, M.~Tanner, C.~Natarajan, R.~Hadfield, S.~Dorenbos, V.~Zwiller, J.~O'Brien, and M.~Thompson, \enquote{{Gallium arsenide (GaAs) quantum photonic waveguide circuits},} 	Opt. Commun.  \textbf{327}, 49--55 (2014).

\bibitem{Guo_LSA_2017}
X. Guo, C. Zou, C. Schuck, H. Jung, R. Cheng, and H. Tang, \enquote{{Parametric down-conversion photon-pair source on a nanophotonic chip},} Light Sci. Appl. \textbf{6}, e16249 (2017).


\bibitem{Harris_PhysRevX_2014}
N. Harris, D. Grassani, A. Simbula, M. Pant, M. Galli, T. Baehr-Jones, M Hochberg, D. Englund, D. Bajoni, and C. Galland, \enquote{{Integrated source of spectrally filtered correlated photons for large-scale quantum photonic systems},} Phys. Rev.  X \textbf{4}, 041047 (2014).

\bibitem{Reck1994}
M.~Reck, A.~Zeilinger, H.~J. Bernstein, and P.~Bertani, \enquote{{Experimental realization of    any discrete unitary operator},} Phys. Rev.  Lett. \textbf{73}, 58--61 (1994).

\bibitem{Clements_Optica_2016}
W.~Clements, P.~Humphreys, B.~Metcalf, W.~S.~Kolthammer, and I.~Walmsley, \enquote{{Optimal      design for universal multiport interferometers},} Optica \textbf{3}, 1460--1465 (2016).

\bibitem{Yariv_JQuantElectron_1973}
A. Yariv, \enquote{{Coupled-Mode Theory for Guided-Wave Optics},} IEEE J. Quant. Elec. \textbf{9}, 919 (1973). 
\bibitem{Soldano_JLightTech_1995}
L. Soldano and E. Pennings. \enquote{{Optical multi-mode interference devices based on self-imaging: principles and applications},} J. Light. Technol. \textbf{13}, 615 (1995). 



\bibitem{Spagnolo_NatureComms_2013}
N. Spagnolo, C. Vitelli, L. Aparo, P. Matalon, F. Sciarrino, A. Crespi, R. Ramponi, and R. Osellame, \enquote{{Three-photon bosonic coalescence in an integrated tritter},} Nat. Commun. \textbf{4}, 1606 (2013).




\bibitem{Orlandi_OL_2013}
P. Orlandi, F. Morichetti, M. Strain, M. Sorel, A. Melloni, and P. Bassi, \enquote{{Tunable silicon photonics directional coupler driven by a transverse temperature gradient},} Opt. Lett. \textbf{38}, 863 (2013).


\bibitem{Sprengers_APL_2011}
J. P. Sprengers, A. Gaggero, D. Sahin, S. Jahanmirinejad, G. Frucci, F. Mattioli, R. Leoni, J. Beetz, M. Lermer, M. Kamp, S. H{\"o}fling, R. Sanjines, and A. Fiore, \enquote{{Waveguide superconducting single-photon detectors for integrated quantum photonic circuits},} Appl. Phys. Lett. \textbf{99}, 181110 (2011).

\bibitem{Bijlani_APL_2013}
B. J. Bijlani, P. Abolghasem, and Amr S. Helmy, \enquote{{Semiconductor optical parametric generators in isotropic semiconductor diode lasers},} Appl. Phys. Lett. \textbf{103}, 091103 (2013).

\bibitem{Boitier_PRL_2014}
F. Boitier, A. Orieux, C. Autebert, A. Lema{\^i}tre, E. Galopin, C. Manquest, C. Sirtori, I. Favero, G. Leo, and S. Ducci, \enquote{{Electrically Injected Photon-Pair Source at Room Temperature},} Phys. Rev.  Lett. \textbf{112}, 183901 (2014).




\bibitem{Zhizhongunpublished} Z. Yan, , {\it  in preparation}. 

\bibitem{Horn_SciRep_2013}
R. Horn, P. Kolenderski, D. Kang, P. Abolghasem, C. Scarcella, A. Frera, A. Tosi, L. Helt, S. Zhukovsky, J. Sipe, G. Weihs, A. Helmy, and T. Jennewein, \enquote{{Inherent polarization entanglement generated from a monolithic semiconductor chip},} Sci. Rep.  \textbf{3}, 2314 (2013).

\bibitem{Abolghasem_OL_2009}
P. Abolghasem, M. Hendrych, X. Shi, J. Torres, and A. Helmy, \enquote{{Bandwidth control of paired photons generated in monolithic Bragg reflection waveguides},} Opt. Lett.  \textbf{34}, 2000 (2009).


\bibitem{Erdogan_OC_1998}
T. Erdogan \enquote{{Optical add-drop multiplexer based on an asymmetric Bragg coupler},} Opt. Commun.  \textbf{157}, 249 (1998).
 
\bibitem{Wang_OL_2007}
T. Wang, C. Chu, and C. Lin, \enquote{{Electro-optically tunable microring resonators on lithium niobate},} Opt. Lett. \textbf{32}, 2777 (2007).
 
\bibitem{Manipatruni_OE_2010}
S. Manipatruni, K. Preston, L. Chen, and M. Lipson, \enquote{{Ultra-low voltage, ultra-small mode volume silicon microring modulator},}  Opt. Express \textbf{18}, 18235 (2010).
 

\bibitem{Miller_PRL_1984}
D. Miller, D. Chemla, T. Damen, A. Gossard, W. Wiegmann, T. Wood, and C. Burrus, \enquote{{Band-edge electroabsorption in quantum well structures: the quantum-confined Stark effect},} Phys. Rev.  Lett.  \textbf{53}, 2173 (1984).


\bibitem{Cariolaro2016}
G.~Cariolaro and G.~Pierobon, \enquote{{Bloch-Messiah reduction of Gaussian unitaries by Takagi factorization},} Phys. Rev.  A \textbf{94}, 062109 (2016).


\bibitem{Adesso2014}
G.~Adesso, S.~Ragy, and A.~R. Lee, \enquote{{Continuous variable quantum
 information: Gaussian states and beyond},} {{Open Syst. Inf. Dyn.}} \textbf{21}, 1440001 (2014).

\bibitem{Ferraro2005}
A.~Ferraro, S.~Olivares, and M.~G.~a. Paris, \emph{{Gaussian states in
  continuous variable quantum information}} (Bibliopoli, Napoli Series on Physics and Astrophysics 2005).
  


\bibitem{Uhlmann_RMP_1976}
A. Uhlmann, \enquote{{The transition probability in the state space of a $\ast$-algebra},} Rep. Math. Phys. \textbf{9}, 273 (1976). 


\bibitem{Harris2017}
N.~C. Harris, G.~R. Steinbrecher, M.~Prabhu, Y.~Lahini, J.~Mower, D.~Bunandar,
  C.~Chen, F.~N.~C. Wong, T.~Baehr-Jones, M.~Hochberg, S.~Lloyd, and
  D.~Englund, \enquote{{Quantum transport simulations in a programmable
  nanophotonic processor},} {Nat. Photonics
}
  \textbf{11}, 447--452 (2017).

\bibitem{Menicucci_PRL_2014}
N. Menicucci, \enquote{{Fault-tolerant measurement-based quantum computing with continuous-variable cluster states},} Phys. Rev.  Lett. \textbf{112}, 120504 (2014).



\bibitem{Fiurasek_PRA_2011}
J. Fiur{\'{a}}{\u{s}}ek, \enquote{{Improving entanglement concentration of Gaussian states by local displacements},} Phys. Rev.  A \textbf{84}, 012335 (2011).

\bibitem{Xiang_NaturePhoton_2010}
G. Xiang, T. Ralph, A. Lund, N. Walk, and G. Pryde, \enquote{{Heralded noiseless linear amplification and distillation of entanglement},} Nat. Photonics
 \textbf{4}, 316 (2010).


\bibitem{Zavatta_NaturePhoton_2010}
A. Zavatta, J. Fiur{\'{a}}{\u{s}}ek, and M. Bellini, \enquote{{A high-fidelity noiseless amplifier for quantum light states},} Nat. Photonics
 \textbf{5}, 52 (2010).


\bibitem{Tan2008}
S.-H. Tan, B.~I. Erkmen, V.~Giovannetti, S.~Guha, S.~Lloyd, L.~Maccone, S.~Pirandola, and J.~H. Shapiro, \enquote{{Quantum Illumination with Gaussian States},} Phys. Rev.  Lett. \textbf{101}, 253601 (2008).

\bibitem{Niset2008}
J.~Niset, U.~L. Andersen, and N.~J. Cerf, \enquote{{Experimentally feasible
  quantum erasure-correcting code for continuousvariables},} Phys. Rev. 
  Lett.  \textbf{101}, 130503 (2008).

\bibitem{Hayden2016}
P.~Hayden, S.~Nezami, G.~Salton, and B.~C. Sanders, \enquote{{Spacetime
  replication of continuous variable quantum information},} N. J. Phys. 
   \textbf{18}, 083043 (2016).

\bibitem{Lu2017a}
D.~Lu, K.~Li, J.~Li, H.~Katiyar, A.~J. Park, G.~Feng, T.~Xin, H.~Li, G.~Long,
  A.~Brodutch, J.~Baugh, B.~Zeng, and R.~Laflamme, \enquote{{Enhancing quantum
  control by bootstrapping a quantum processor of 12 qubits},} npj Quantum
  Information \textbf{3}, 45 (2017).





\bibitem{Dive2017}
B.~Dive, A.~Pitchford, F.~Mintert, and D.~Burgarth, \enquote{{In situ upgrade
  of quantum simulators to universal computers},}  https://arxiv.org/abs/1701.01723 (2017).
  
\bibitem{Yokoyama_NaturePhoton_2013}
S. Yokoyama, R. Ukai, S. Armstrong, C. Sornphiphatphong, T. Kaji, S. Suzuki, J. Yoshikawa, D. Yonezawa, N. Menicucci, and A. Furusawa, \enquote{{Ultra-large-scale continuous-variable cluster states multiplexed in the time domain},} Nat. Photonics
 \textbf{7}, 982 (2013).

\bibitem{Yoshikawa_APLPhoton_2016}
J. Yoshikawa, S. Yokoyama, T. Kaji, C. Sornphiphatphong, Y. Shiozawa, K. Makino, and A. Furusawa, \enquote{{Generation of one-million-mode continuous-variable cluster state by unlimited time-domain multiplexing},} APL Photonics \textbf{1}, 060801 (2016).






\end{thebibliography}
\end{document}